# Optimal Time-Trial Bicycle Racing with Headwinds and Tailwinds

A. Brad Anton
Cornell University
aba6@cornell.edu

## Abstract

Many time-trial and triathlon bicycle races take place on relatively flat, closed-circuit courses. In the absence of hills, riding-speed is limited almost solely by aerodynamic drag; consequently, winds can have a big effect on elapsed times. I analyze the special case of a straight out-and-back race in a steady wind, assuming the rider has a given total amount of energy to expend and can choose only two speeds – the aided speed with tailwind and the hindered speed into headwind. In this ideal circumstance the problem of choosing optimal riding speeds reduces to a constrained nonlinear optimization that can be solved with elementary calculus. My analysis reveals a practical "rule of thumb" that can be used more generally to choose optimal riding speeds for time-trial racing on closed-circuit courses in the presence of headwinds and tailwinds.

**Introduction**

Many time-trial bicycle races take place on relatively flat, closed-circuit courses, and the bicycling portions of triathlons are often on flat ground near the shores of lakes or oceans where the swimming events are held. When there are no grand hills to climb, winds can have a big impact on elapsed times. Once you have a good bike, a good riding position, a good pedaling stroke, and good fitness, the next thing you may need for succeeding in a time-trial race is a clever strategy for managing your effort in a steady wind.

The question I address here is how you choose and vary your riding speed in windy conditions to optimize your overall speed. I seek a "rule of thumb" that can be implemented on-the-fly, in your head, without requiring sophisticated calculations, pre-race measurements in a laboratory, or on-course communications with a coach. My analysis assumes the simplest imaginable scenario: a flat, straight-line, out-and-back course with a steady wind, with aerodynamic drag being the only force limiting, and with only two speeds to choose, the aided speed with tailwind and the hindered speed with headwind. Though the straight out-and-back model is restrictive, the conclusions that derive from it are more general, because a flat, closed-circuit course of most any configuration can be decomposed into parallel, out-and-back segments with compensating headwinds and tailwinds; the results herein apply to all such segments independently.

Optimizing your overall speed in this race involves solving two problems[1].

One is a physics problem that describes the interaction of your bicycle and your body with the road and the air. It involves mathematically rigorous laws of nature that apply to every rider in every circumstance: we know, for example, exactly how to calculate power as a function of speed and drag force, and we know that resistance from aerodynamic drag dominates all other drag forces in high-speed cycling (e.g., >15 mph)[2,3]. This is why modern time-trail bicycles have airfoil-shaped tubes and hidden control cables, and why their riders crouch in an awkward position and often wear skin suits and boat-tailed safety helmets.

The other problem is human physiology, which describes how your body generates pedaling-power from the food you eat and the air you breathe[1,3]. A human body operates like a complicated biofuel-cell, and the quantitative measures of its performance are numbers like $VO_{2max}$ and lactate threshold that defy first-principles mathematical modeling[3]. We know that good nutrition, a good training regimen, and good rest matter a lot (and, lamentably, that performance-enhancing drugs can matter the most), but there are no biophysical equations that predict their quantitative effects on cycling performance for everyone.

For the sake of the arguments that follow, I will disguise the uncertainty of human physiology in the mathematical certainty of physics by assuming a cyclist is simply a machine that can deliver pedaling-power without limitations, but has a fixed inventory of energy (i.e. thermodynamic work) to expend in the race. Simplicity is a virtue here, because it reduces the optimal-speed problem to a constrained nonlinear optimization that can be solved with elementary calculus. The solution may not account for your particular physiological capabilities or limitations, but it reveals what you *should* do if you *can*, and it rationalizes *why* you should do it.

**Develop the Model**

Imagine you ride out and back on a flat road of length $= \frac{\ell}{2}$, a total distance $= \ell$, with a steady wind at speed $= w$ into your face in one direction and at your back the other direction (or an angled crosswind that resolves into a headwind and tailwind). What speeds should you ride out and back to give you the highest possible average speed for the round trip?

First, let's establish a wind-free calibration for comparison. Pick a windless day and ride the time-trial course out and back as fast as you can. Let $v_0$ be the constant speed you have ridden; it is the target-speed you would endeavor to ride on that course in a time-trial race if the conditions were ideal, or if you are a triathlete, it is the constant speed you would ride over that distance on a wind-free day while conserving enough energy for swimming and running. The total time for the out-and-back ride is $t_0 = \frac{\ell}{v_0}$. Assume the only resistance is aerodynamic drag, in which case the steady drag force is $F_0 = \frac{C_D A \rho v_0^2}{2}$, where $C_D =$ drag coefficient, $A =$ frontal area, and $\rho =$ air density[2,3]. Your steady power output for the calibration ride is

[1] $\qquad P_0 = F_0 v_0 = \frac{C_D A \rho v_0^3}{2}$.

(Measurements for skilled cyclists riding modern aerodynamic bicycles on flat ground in wind-free conditions show power outputs around $P_0 \cong 175$ Watts for $v_0 \cong 20$ mph and $P_0 \cong 600$ Watts for $v_0 \cong 30$ mph, for example[1,3].) The total energy you expend for the round trip is

[2] $\qquad W_0 = F_0 \ell = \frac{C_D A \rho v_0^2 \ell}{2}$.

Let's assume this $W_0$ quantifies the maximum sustained effort you can deliver for an out-and-back ride on this particular course, windy or not. If you were a car, $W_0$ would be a full tank of fuel.

Now do the ride again, but this time there is a steady headwind in one direction and an equal tailwind in the other. Let $v_+$ be the higher speed with the wind at your back, and let $v_-$ be the lower speed with the wind in your face. The average speed for the round trip is

[3] $\qquad v = \frac{\ell}{t} = \frac{\ell}{\frac{\ell}{2v_+} + \frac{\ell}{2v_-}} = \frac{2v_+ v_-}{v_+ + v_-}$.

The aerodynamic drag forces are $F_+ = \frac{C_D A \rho (v_+ - w)^2}{2}$ with tailwind and $F_- = \frac{C_D A \rho (v_- + w)^2}{2}$ into headwind, and the corresponding power outputs are respectively. The total energy expended for the round trip is which we will assume is $W_0$, exactly the same as the energy you expended for your maximum-effort, wind-free calibration ride at speed $= v_0$. We seek the choices of $v_+$ and $v_-$ that maximize $v$ of equation **[3]**, subject to the constraint that the wind-affected $W$ of equation **[5]** is equal to the wind-free $W_0$ of equation **[2]**. We will assume the parameters $C_D$, $A$, $\rho$, $\ell$, and $w$ are immutable constants in this situation, so we can isolate the optimally constrained relationships among $v_+$, $v_-$, and $w$.

[4] $$P_+ = \frac{C_D A \rho v_+ (v_+ - w)^2}{2} \text{ and } P_- = \frac{C_D A \rho v_- (v_- + w)^2}{2},$$

[5] $$W = \frac{F_+ \ell}{2} + \frac{F_- \ell}{2} = \frac{C_D A \rho \ell}{4}\left[(v_+ - w)^2 + (v_- + w)^2\right],$$

One can envisage several strategies for optimizing the average wind-affected speed. It is easy to prove that riding at a constant speed maximizes speed and minimizes total energy consumption when there are no external forces. Perhaps, then, your best strategy for a wind-affected round-trip would be to ride at *equal speeds*, going against the wind with whatever power is necessary to ensure that the remainder of your energy-inventory will be just enough to cover the other, wind-aided leg at exactly the same speed. Or perhaps it's better to deliver *equal power outputs*, keeping yourself at the "sweet spot" of pedal-rpm, heart-rate, and breathing while changing gears to ride at whatever out-and-back speeds happen. (This feels best to me.) Or maybe the best strategy is actually something in between, where you work a bit harder while riding into the wind, maybe even approach the "red-line zone" of high heart rate, and you relax a bit while riding with the wind, so as to optimally decrease (but not completely erase) the difference between your wind-aided and wind-hindered speeds. Let's call this the *optimal speeds* scenario.

Before we consider any of these strategies, let's convert the problem into a simpler, dimensionless form. Let $x = \frac{v_+}{v_0}$ be the scaled wind-aided speed ($x > 1$); let $y = \frac{v_-}{v_0}$ be the scaled wind-hindered speed ($y < 1$); let $\alpha = \frac{w}{v_0}$ be the scaled wind speed ($\alpha < 1$), and let $f = \frac{v}{v_0}$ be the scaled average speed. In terms of these, the function we seek to maximize (equation **[3]**) becomes

[6] $$f(x, y) = \frac{2xy}{x + y},$$

and the constraint that we only have a fixed, total amount of energy to use (equation **[5]**) becomes

[7] $$(x - \alpha)^2 + (y + \alpha)^2 = 2.$$

The scaled power outputs are

[8] $$p_x = \frac{P_+}{P_0} = x(x - \alpha)^2 \text{ and } p_y = \frac{P_-}{P_0} = y(y + \alpha)^2$$

for the wind-aided and wind-hindered legs, respectively, where $P_0$ is the baseline power (equation **[1]**). Solving these equations for different scenarios – constant speeds, constant power outputs, or optimal speeds – gives different functions $x(\alpha)$ and $y(\alpha)$. If we convert them back to the original, dimensioned variables – road speeds $v_+$, $v_-$ and wind speed $w$ – we will have a recipe for adjusting the headwind and tailwind riding speeds to meet the stated objective while conserving a fixed, total amount of energy.

**Strategy #1 - Equal Speeds**

Let's solve the easiest problem first: *equal speeds*, so $v_+ = v_-$, or in scaled variables $x = y = f$. Solve equation **[7]** with $x = y$ to find for the scaled speed, and substitute these into **[8]** to find the scaled power outputs $p_x$ and $p_y$. Figure 1, below, shows a plot of the results.

**[9]** $\quad x(\alpha) = y(\alpha) = \sqrt{1-\alpha^2}$

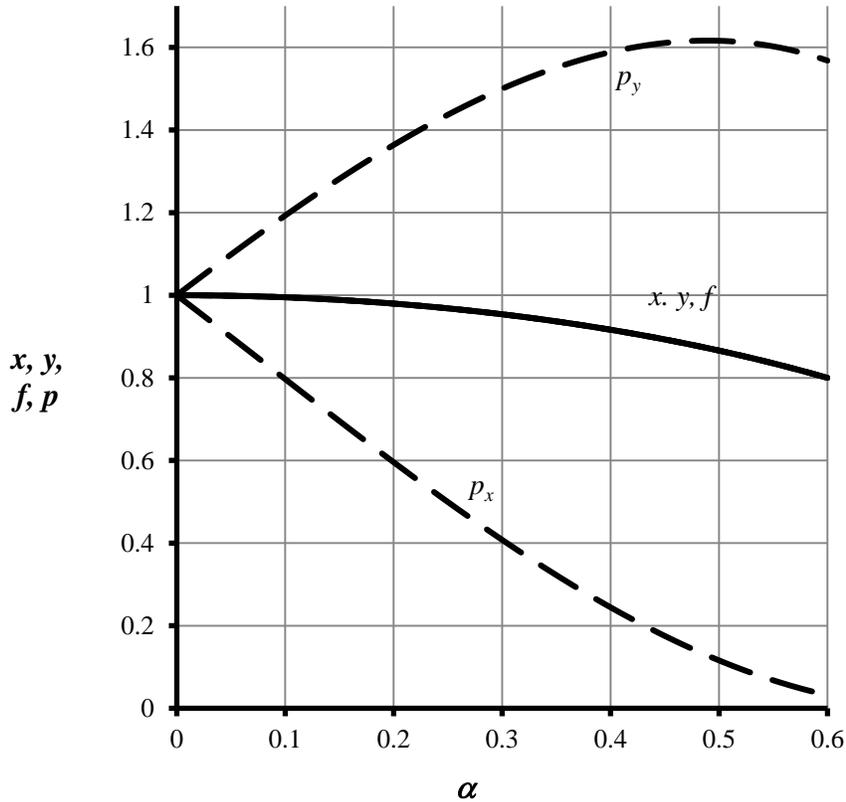

**Figure 1 – Equal Speeds.** Scaled speed $x = y = \dfrac{v}{v_0} = f$ (solid line) and power outputs $p_x = \dfrac{P_+}{P_0}$ and $p_y = \dfrac{P_-}{P_0}$ (dashed lines) as functions of the scaled headwind/tailwind speed $\alpha = \dfrac{w}{v_0}$.

Notice that the overall speed decreases unavoidably from the wind-free speed $f = 1$ as $\alpha$ increases. Since the aerodynamic drag force scales quadratically with $(v \pm w)$, one will always consume more energy riding into the wind than one gains back riding with the wind on a closed-circuit course, which ensures a lower overall speed when only a fixed amount of energy is available for the round trip.

Notice also what happens when $\alpha \cong 0.5$, i.e. the head- and tailwind speed is about half of the target riding speed. The power output $p_y$ for riding into the wind grows to ~1.6x the steady power output for the wind-free calibration ride, which constitutes a huge, likely unsustainable effort, particularly in a triathlon. The cost for putting out such a Herculean effort against the wind is there is no energy left for the wind-aided leg of the trip: $p_x$ drops nearly to zero. You are effectively killing yourself against the wind and nearly coasting with the wind to maintain equal speeds out and back, which hardly seems optimal.

**Strategy #2 - Equal Power Outputs**

Now let's investigate the scenario with *equal power outputs*. In this case $p_x = p_y$ in equation **[8]**, or

[10] $\quad x(x-\alpha)^2 = y(y+\alpha)^2.$

This must be solved simultaneously with equation **[7]**, the scaled version of total energy constraint. They are nonlinearly coupled, hence strenuous to solve in closed form, but they can be decoupled and solved approximately with a power-series expansion in $\alpha$. Let $x(\alpha) \cong x_0 + \alpha x_1 + \alpha^2 x_2$ and $y(\alpha) \cong y_0 + \alpha y_1 + \alpha^2 y_2$. Substitute these into **[7]** and **[10]**, and collect equations by orders in $\alpha$. One finds:

At $O(1)$ (="order of one"):
$$x_0^2 + y_0^2 = 2 ;$$
$$x_0^3 = y_0^3$$
$$\to x_0 = 1 \text{ and } y_0 = 1.$$

At $O(\alpha)$:
$$x_1 + y_1 = 0 ;$$
$$3x_1 - 2 = 2 + 3y_1$$
$$\to x_1 = \frac{2}{3} \text{ and } y_1 = -\frac{2}{3}.$$

And at $O(\alpha^2)$:
$$x_2 - y_2 = 0 ;$$
$$-x_1 = \frac{1}{9} + y_1$$
$$\to x_2 = -\frac{1}{18} \text{ and } y_2 = -\frac{1}{18}.$$

Then

[11] $\quad x(\alpha) \cong 1 + \dfrac{2\alpha}{3} - \dfrac{\alpha^2}{18}$ and $y(\alpha) \cong 1 - \dfrac{2\alpha}{3} - \dfrac{\alpha^2}{18}.$

These series expansions for $x(\alpha)$ and $y(\alpha)$, though obtained with minimal effort, are remarkably accurate. For example, an exact numerical solution of the coupled nonlinear equations **[7]** and **[10]** gives $x = 1.373$ and $y = 0.585$ for $\alpha = 0.60$, whereas the approximations **[11]** give $x = 1.380$ and $y = 0.580$, agreeing with the exact answers to within 1%. The functions $f$, $p_x$, and $p_y$ derived from them are similarly accurate, as are comparisons among these functions for different speed strategies, coming later in this analysis.

Figure 2 shows a plot of these functions, along with the scaled average speed $f$ from equation **[6]** and the scaled power output $p_x = p_y$ from equation **[8]**.

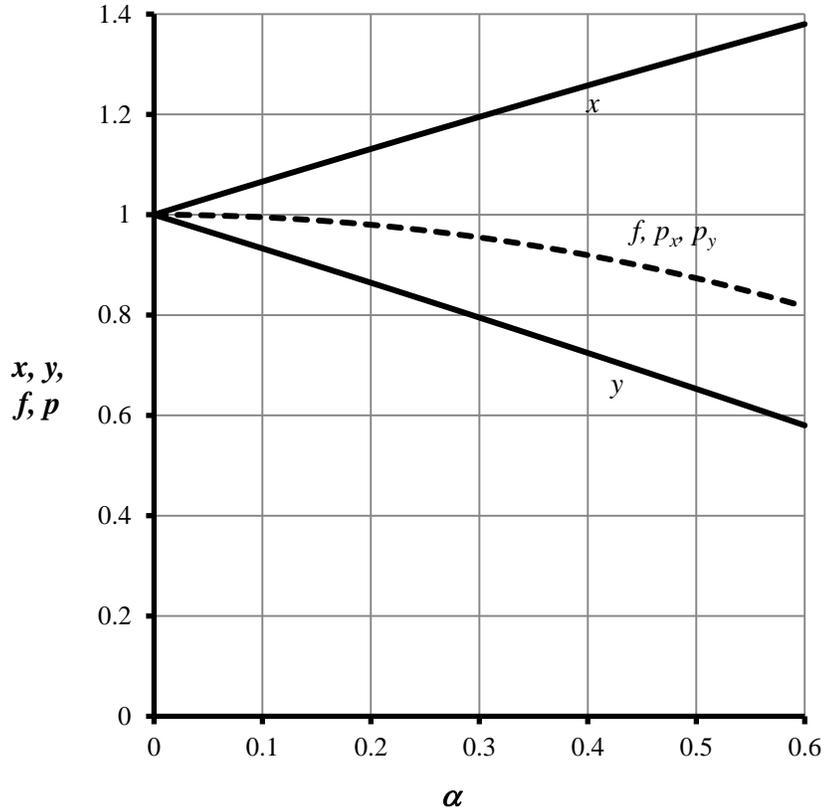

**Figure 2 – Equal Power Outputs.** Scaled out and back speeds $x = \dfrac{v_+}{v_0}$ and $y = \dfrac{v_-}{v_0}$ (solid lines) and average speed $f = \dfrac{v}{v_0}$ (dashed line) as functions of the scaled headwind/tailwind speed $\alpha = \dfrac{w}{v_0}$. The curve for the power outputs $p_x = \dfrac{P_+}{P_0} = p_y = \dfrac{P_-}{P_0}$ is virtually coincident with $f$, so an extra line is not shown.

One sees that the difference between the scaled, wind-aided speed $x$ and the wind-hindered speed $y$ increases as the wind speed $\alpha$ increases. This causes the average speed $f$ to decrease, which increases the total time on course. The steady power output $p_x = p_y$ decreases because a fixed amount of total energy is being consumed over a longer time.

**Strategy #3 - Optimal Speeds**

Now let's see if we can find *optimal speeds* and power outputs for the headwind and tailwind directions that yield the highest possible overall speed. We must find new functions $x(\alpha)$ and $y(\alpha)$ that maximize equation **[6]** for $f(x, y)$ while meeting the total energy constraint, equation **[7]**. This is an optimization with an equality constraint; consequently, the venerable "Method of Lagrange Multipliers" applies. The Lagrange function in this case combines **[6]** and **[7]** as follows:

[12] $$\Lambda(x, y, \lambda) = \frac{2xy}{x+y} + \lambda\left[(x-\alpha)^2 + (y+\alpha)^2 - 2\right],$$

where $\lambda$ is the Lagrange multiplier. The optimal choices of $x$ and $y$ are realized by solving the following three equations to eliminate $\lambda$ and recover $x$ and $y$ as parametric functions of $\alpha$:

[13A] $$\left(\frac{\partial \Lambda}{\partial x}\right)_{y,\lambda,\alpha} = 0 = \left(\frac{y}{x+y}\right)^2 + \lambda(x-\alpha);$$

[13B] $$\left(\frac{\partial \Lambda}{\partial y}\right)_{x,\lambda,\alpha} = 0 = \left(\frac{x}{x+y}\right)^2 + \lambda(y+\alpha); \text{ and}$$

[13C] $$\left(\frac{\partial \Lambda}{\partial \lambda}\right)_{x,y,\alpha} = 0 = (x-\alpha)^2 + (y+\alpha)^2 - 2.$$

A few moves to eliminate $\lambda$ among **[13A-C]** returns this cubic relationship

[13D] $$x^2(x-\alpha) = y^2(y+\alpha).$$

The functions $x(\alpha)$ and $y(\alpha)$ we seek in this case are the simultaneous solutions to equations **[13C]** and **[13D]**, which are again nonlinearly coupled and strenuous to solve in closed form. Using the power-series method exactly as before gives

[14] $$x(\alpha) \cong 1 + \frac{\alpha}{3} - \frac{\alpha^2}{9} \text{ and } y(\alpha) \cong 1 - \frac{\alpha}{3} - \frac{\alpha^2}{3}.$$

Figure 3 shows a plot of these functions, along with the scaled average speed $f$ from equation **[6]** and the scaled power outputs $p_x$ and $p_y$ from equation **[8]**.

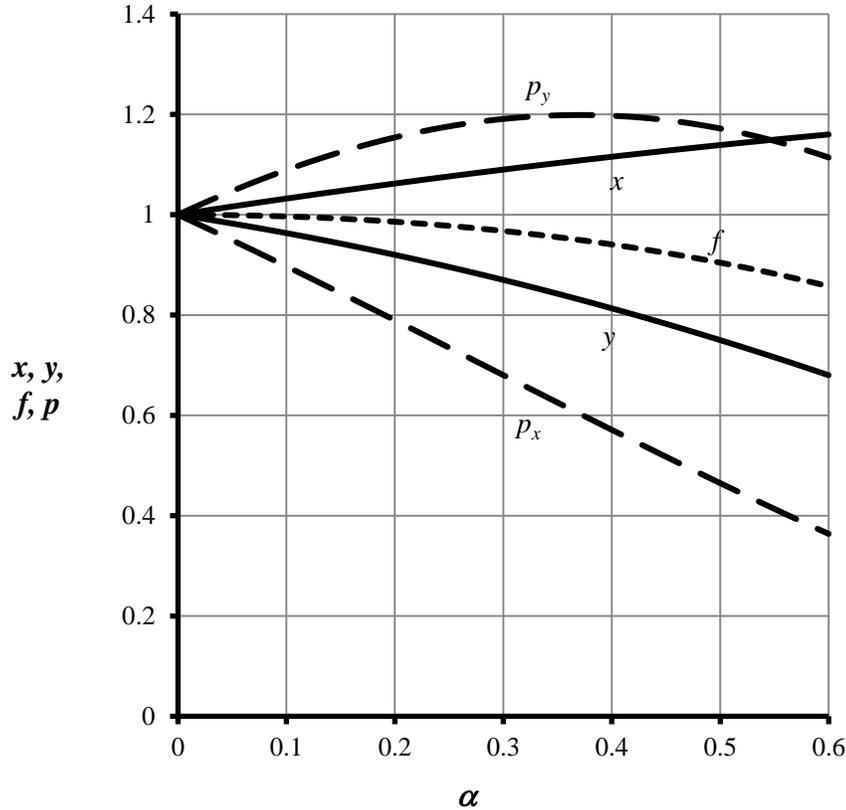

**Figure 3 – Optimal Speeds.** Scaled speeds $x = \frac{v_+}{v_0}$ and $y = \frac{v_-}{v_0}$ (solid lines), average speed $f = \frac{v}{v_0}$ (short-dashed line), and power outputs $p_x = \frac{P_+}{P_0}$ and $p_y = \frac{P_-}{P_0}$ (long-dashed lines) as functions of the scaled headwind/tailwind speed $\alpha = \frac{w}{v_0}$.

Notice that the power output against the wind $p_y$ is still substantially greater than the power output with the wind $p_x$, but this optimal difference in power outputs is much less than the extreme difference we found for the equal-speeds scenario of Figure 1. In this case it decreases the difference between the wind-aided and wind-hindered speeds $x$ and $y$ just the right amount to optimally increase the overall speed $f$ for the given total amount of energy expended. The power output $p_y$ against the wind peaks at ~1.2x the calibration power when the headwind speed is 0.3-0.4x the calibration speed, which constitutes a strenuous effort but is much more likely sustainable than the ~1.6x required in the equal-speeds scenario.

**Comparison of Strategies**

Figure 4 compares the average speed curves from Figures 1 through 3. Notice that the equal speeds and equal power recipes give almost the same average speed curve, but the optimal speeds curve is significantly better than either, particularly at high wind speeds. For example, the percentage improvement at $\alpha \cong 0.5$ is approximately $\left(\frac{.905-.87}{.87}\right)100\% \cong 4.0\%$. This is worth about 2.4 min in a one hour race, which could be enough to make a dramatic difference in finishing position. And remember, this improvement has been realized while consuming the same total amount of pedaling energy, so apparently for free.

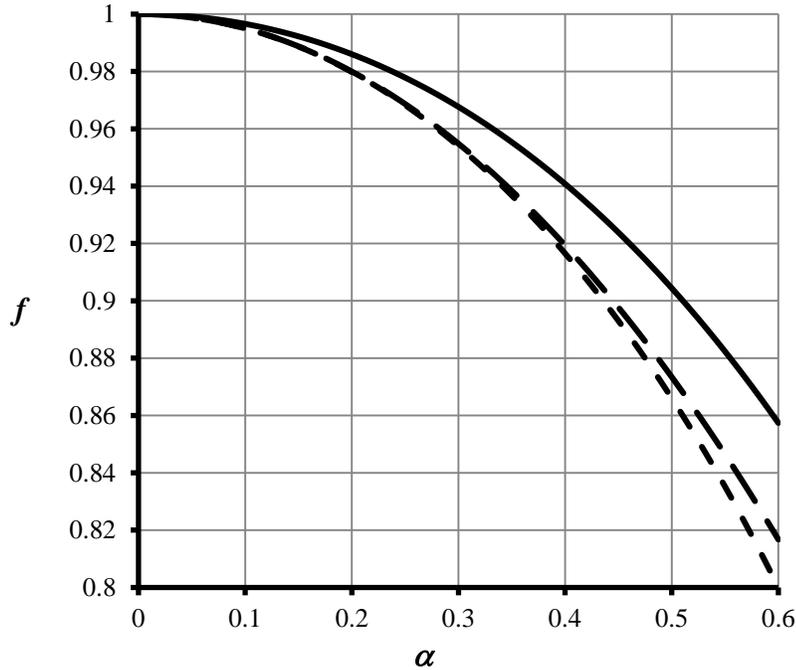

**Figure 4 – Comparison.** Average speed $f = \frac{v}{v_0}$ vs. scaled headwind/tailwind speed $\alpha = \frac{w}{v_0}$ for *equal speeds* (short-dashed line, cf. Figure 1), *equal power outputs* (long-dashed line, cf. Figure 2), and *optimal speeds* (solid line, cf. Figure 3).

**A Realistic Example**

Let's see what these imply for a realistic example[4]. Begin by gathering up the analytical results. Refer to Table 1, below.

**Table 1 – Summary of Results.**

| Strategy | Speeds | Power Outputs | Average Speed |
|---|---|---|---|
| **Optimal Speeds** | $x = \dfrac{v_+}{v_0} = 1 + \dfrac{\alpha}{3} - \dfrac{\alpha^2}{9}$<br>$y = \dfrac{v_-}{v_0} = 1 - \dfrac{\alpha}{3} - \dfrac{\alpha^2}{3}$ | $p_x = \dfrac{P_+}{P_0} = x(x-\alpha)^2$<br>$p_y = \dfrac{P_-}{P_0} = y(y+\alpha)^2$ | $f = \dfrac{v}{v_0} = \dfrac{2xy}{x+y}$ |
| **Equal Power Outputs** | $x = \dfrac{v_+}{v_0} = 1 + \dfrac{2\alpha}{3} - \dfrac{\alpha^2}{18}$<br>$y = \dfrac{v_-}{v_0} = 1 - \dfrac{2\alpha}{3} - \dfrac{\alpha^2}{18}$ | | |
| **Equal Speeds** | $x = y = \dfrac{v}{v_0} = \sqrt{1-\alpha^2}$ | | |

Imagine a flat time-trial, 12 miles out due north, 12 miles back due south, 24 miles total, and imagine there is a steady 12 mph wind coming from the northeast, so at a 45° angle to the course. Consequently, there is a headwind-speed of $w = 12\cos 45° \cong 8.5$ mph on the out-leg, and a tailwind-speed of 8.5 mph on the back-leg.

Consider a strong competitor who would expect to ride the course at $v_0 = 25$ mph if he delivered an all-out effort with no wind. Assume his on-bike Watt-meter reads 350 W when he rides 25 mph on a flat road with no wind. When the headwind/tailwind speed is 8.5 mph, $\alpha = \dfrac{8.5}{25} = 0.34$ for him.

First use the *optimal speeds* formulas in Table 1 to quantify his best-possible performance. Calculate:
- $x = 1 + \dfrac{0.34}{3} - \dfrac{0.34^2}{9} = 1.10$;
- $y = 1 - \dfrac{0.34}{3} - \dfrac{0.34^2}{3} = 0.85$;
- $p_x = 1.10x(1.10 - 0.34)^2 = 0.64$;
- $p_y = 0.85x(0.85 + 0.34)^2 = 1.20$; and
- $f = \dfrac{2x1.10x0.85}{1.10 + 0.85} = 0.96$.

Now convert the results to dimensioned variables:
- He should ride with the wind at $v_+ = v_0 x = 25x1.1 = 27.5$ mph, and he should expect to see $P_+ = P_0 p_x = 350x0.64 = 220\text{W}$ on his Watt-meter while he's doing it.

- He should ride into the wind at $v_- = v_0 y = 25 \times 0.85 = 21.3$ mph, and his Watt-meter will show $P_- = P_0 p_y = 350 \times 1.20 = 420$ W while he's doing it.
- His overall average speed will be $v = v_0 f = 25 \times 0.96 = 24.0$ mph.
- He will complete the round trip in $t = \frac{\ell}{v} = \frac{24}{24} = 1.00$ hr.

Now repeat the calculation for the *equal power outputs* formulas in Table 1. The dimensionless results are: $x = 1.22$; $y = 0.77$; $p_x = p_y = 0.94$; and $f = 0.94$. If he follows this recipe, he will ride with the wind at $v_+ = 30.5$ mph; he will ride against the wind at $v_- = 19.3$ mph; his power output will be $P_+ = P_- = 330$ W in both directions; his average speed will be $v = 23.6$ mph; and he will compete the round-trip in $t = 1.02$ hr, about a minute slower than if he followed the *optimal speeds* recipe, above. Remember – this recipe consumed exactly the same total amount of energy, just delivered it at a less profitable rate.

Finally, repeat the calculation for the *equal speeds* formulas in Table 1. The dimensionless results are: $x = y = f = 0.94$; $p_x = 0.34$; and $p_y = 1.54$. If he follows this recipe, he will ride both legs at $v = 23.5$ mph; his power output will be a pleasant $P_+ = 120$ W with the wind but a brutal $P_- = 540$ W against the wind; and he will complete the round trip in little over $t = 1.02$ hr, only a fraction of a minute slower than if he followed the *equal power outputs* recipe, but more than a minute slower than if he followed the *optimal speeds* recipe.

If you do the calculations for a slower rider on the same route with the same wind speed (e.g., a triathlete who must ride slower because cycling is only one-third of his race), the effect of wind on average speed is larger, as are the fractional differences among the three recipes. This is because the same value of $w$ with a lower value of $v_0$ implies a larger value of $\alpha = \frac{w}{v_0}$, so the values of all the dimensionless functions come from points farther to the right on the curves of Figures 1, 2, and 3, where the $f$-curves diverge from each other. This means the slower you ride, the more important it is to have a good strategy for fighting the wind.

**A "Rule of Thumb" for Racing in the Wind**

Is there a simple "rule of thumb" that can be extracted from this analysis? Yes, I think so….

First, recognize that the equal power outputs recipe, which would have you maintain the same pedal cadence and heart rate in headwind or tailwind, may feel optimal, but it actually isn't. In fact, it is only barely faster than suffering the punishing swing in power-output that would be required to maintain equal out-and-back speeds. Your overall speed (and your finishing position, of course) will benefit from expending some extra energy when the wind is in your face and conserving some energy when the wind is at your back[5], but *not too much*, because going too far slows you down again as you approach the equal-speeds scenario.

Now look carefully at Figure 3. The lines for the optimal speed functions $x(\alpha)$ and $y(\alpha)$ are subtly curved, but they can be approximated well with straight lines over the range $0 < \alpha < 0.6$. One can say to sufficient accuracy for on-the-fly decisions that $x \cong 1 + \frac{\alpha}{4}$ and $y \cong 1 - \frac{\alpha}{2}$. These approximate linear functions predict $x \cong 1.10$ and $y \cong 0.80$ at $\alpha = 0.4$, for example, which are within 2% of the optimal choices. Here is the rule of thumb we seek!

But these are scaled, dimensionless functions that convert to real road speeds only if you know your target speed $v_0$ for a wind-free ride on the race course, since it is the baseline on which your strategy is based, and you know the angle-resolved head/tailwind speed $w$ everywhere on the course on the day of your race. Experience, research, and planning are vital to gathering these data accurately and using them effectively.

Choose a target-speed $= v_0$ for the closed-circuit course you are riding and estimate the angle-resolved wind-speed $= w$ for every leg of the course. Endeavor to ride at $v \cong v_0 + \frac{w}{4}$ when the wind is at your back and at $v \cong v_0 - \frac{w}{2}$ when the wind is at your face. Vary your speed continuously as your angle with the wind or the wind-speed itself changes, anticipating (and hoping) that the winnings from compensating out-and-back segments of the closed course will accumulate to your advantage. If you are vigilant (and maybe a bit lucky), you might win enough time or conserve enough energy with this trick to catch or drop other riders who have a less profitable strategy for fighting the wind. Good luck!